\documentclass[12pt]{amsart}
\usepackage{amsmath}
\usepackage{amsthm}
\usepackage{amsfonts}
\usepackage{graphicx}
\usepackage{amssymb}% for \sphericalangle
\newcommand{\arcangle}{\mathord{<\mspace{-9mu}\mathrel{)}\mspace{2mu}}}

\begin{document}

\title[Another View on the Shape Equation for Strings]
{Another View on the Shape Equation for Strings}
\author[Jens Hoppe ]{JENS HOPPE}
%\date{June 14, 2020}
\address{Braunschweig University, Germany}
\email{jens.r.hoppe@gmail.com}

\begin{abstract}
The question how an $M$-dimensional extended object must be shaped
so that a rigid motion gives an $M$-brane solution ($M+1$ dimensional
timelike zero mean curvature surface) in $M+2$ dimensional Minkowski space
is discussed for closed strings
\end{abstract}

\maketitle
%%%%%%%%%%%%%%%%%%%%%%%%%%%%%%%%%%%%%%%%%%%%%%%%%%%%%%%%%%%%%%%%%%%%%%%%%%%%%
%PAGE 1/1 PAGE 1/1 PAGE 1/1 PAGE 1/1 PAGE 1/1 PAGE 1/1 PAGE 1/1 PAGE 1/1 PAGE 1/1
%%%%%%%%%%%%%%%%%%%%%%%%%%%%%%%%%%%%%%%%%%%%%%%%%%%%%%%%%%%%%%%%%%%%%%%%%%%%%
In \cite{1}, an equation was derived for the shape that a planar curve must have to via rigid rotation generate a time-like worldsheet of zero mean curvature in $\mathbb R^{1,2}$. It was found that the shape must  contain singularities (whose worldlines, i.e. propagations, consequently are helices), just as all closed string solutions in 3 dimensional Minkowski-space(assuming they start out regular) develop singularities in finite time. A thorough mathematical treatment was given in \cite{2} (see also \cite{8}), and (for the vast cosmic string literature see e.g. \cite{3,4} and references therein) concerning earliest related references I would like to point out \cite{5}\footnote{weirdly, although quoting it in \cite{1} (Klaus Happle, who was writing his Ph.D thesis on classical strings at that time, must have mentioned or sent \cite{5} to me shortly before the submission of \cite{1}) I did not closely look at it until a few days ago - discovering that their findings are in yet another way strongly related to the present article (they explicitly mention epicycloid/hypocycloid solutions)}. While \cite{1} does give the most prominent solutions of the shape-equation (and mentions the standard form of general string solutions) it does not deduce (nor did \cite{6}) the general solution of the shape equation. The solution on p.21 of \cite{7}\footnote{containing several typos (it should be $r', w, \gamma_0^2,\cdot$ instead of $r,\omega, \gamma_0, +$)}, on the other hand  does not explicitly mention the regime $\gamma_0^2 < 1 (w \in [\gamma_0^2, 1])$, respectively when it does (II. 64), with a typo ($>1 \ldots$ instead of $<1\ldots$), although the derivation would be just the same - the constants in the final formula (84) can simply be taken to be arbitrary. In this note I would like to present an alternative derivation, which stresses the \textit{curvature} of the planar curve and, apart from providing many additional aspects of the problem gives a conformal parameterization of all the world-sheet solutions.\\\\
%%%%%%%%%%%%%%%%%%%%%%%%%%%%%%%%%%%%%%%%%%%%%%%%%%%%%%%%%%%%%%%%%%%%%%%%%%%%%
%PAGE 2 PAGE 2 PAGE 2 PAGE 2 PAGE 2 PAGE 2 PAGE 2 PAGE 2 PAGE 2 PAGE 2 PAGE 2
%%%%%%%%%%%%%%%%%%%%%%%%%%%%%%%%%%%%%%%%%%%%%%%%%%%%%%%%%%%%%%%%%%%%%%%%%%%%%
Following \cite{1}, time-like zero-mean-curvature surfaces in $\mathbb R^{1,2}$ of the form
\begin{equation}\label{eq1}
\left( x^{\mu}\right)  =
\begin{pmatrix}
t\\
\vec{x}(t,\varphi) = e^{f(t)A}\vec{u}(\varphi)
\end{pmatrix}
\end{equation}
with\footnote{note the typo in (54)\cite{1}} $ A = \big(\begin{smallmatrix} 0 & -1 \\ 1 & 0 \end{smallmatrix}\big)$, i.e. obtained by rigidly rotating a planar curve $\vec{u}(\varphi)$ must satisfy the `shape-equation'\footnote{any classical geometer would probably immediately recognize what its solution-curves are, but the route I will take may independently be of some interest}
\begin{equation}\label{eq2}
\omega^2r^2(1+\gamma sin^2 \phi) = 1
\end{equation}
where $\phi := \arcangle(\vec{u},\vec{u}')$, $\gamma = \frac{1}{\gamma^2_0} - 1 > -1$, and $f(t) = \omega t$ ($+f_0$, i.e. the planar curve necessarily rotated with constant angular velocity), all following from, the $\mu =0$ part of the minimal surface equation
\begin{equation}\label{eq3}
\dfrac{1}{\sqrt{G}}\partial_{\alpha}(\sqrt{G}G^{\alpha \beta} \partial_{\beta}x^{\mu}) = 0,
\end{equation}
$G := -det(G_{\alpha \beta} := \partial_{\alpha}x^{\mu}\partial_{\beta}x^{\nu}\eta_{\mu\nu})$, $\eta_{\mu\nu} = diag(1,-1,-1)$ which (after first concluding the allowed $f(t)$ from it) reads
\begin{equation}\label{eq4}
\dfrac{\partial}{\partial \varphi} \underbrace{\left(\dfrac{\omega r sin \phi}{\sqrt{1-\omega^2 r^2 cos^2 \phi}}\right)}_{=: \gamma_0}  = 0,
\end{equation}
%%%%%%%%%%%%%%%%%%%%%%%%%%%%%%%%%%%%%%%%%%%%%%%%%%%%%%%%%%%%%%%%%%%%%%%%%%%%%
%PAGE 3 PAGE 3 PAGE 3 PAGE 3 PAGE 3 PAGE 3 PAGE 3 PAGE 3 PAGE 3 PAGE 3 PAGE 3
%%%%%%%%%%%%%%%%%%%%%%%%%%%%%%%%%%%%%%%%%%%%%%%%%%%%%%%%%%%%%%%%%%%%%%%%%%%%%
Solutions of (\ref{eq2}) clearly fall into 2 distinctive groups\footnote{typo in II.64}: $\gamma > 0 \; (\gamma_0^2 < 1, w := \omega^2 r^2 \in [\gamma_0^2,1])$ and $\gamma < 0 \; (\gamma_0^2 > 1, \omega^2 r^2 \in [\gamma_0^2,1])$,
while in both cases (\ref{eq2}) immediately implies that the curve can not be regular (as according to (\ref{eq2}) $sin^2 \phi$ can not be maximal at $r_{\text{Max}}$ if $\gamma > 0$, resp. at $r_{\text{Min}}$ if $\gamma < 0$ - cp.\cite{1}, but note that the first word in the second line after (58) should be deleted). Instead of solving for the angle $\theta$ in $\vec{u} = r(\varphi)\big(
\begin{smallmatrix}
cos \theta(\varphi)\\
sin \theta(\varphi)
\end{smallmatrix}\big)
$
as a function of $r$ (done on p.21 of \cite{6} for $\gamma < 0$; that derivation goes through identically when $\gamma > 0$, despite of the geometrically very different result), let us solve for the shape by going to local arclength parametrization (i.e. starting at a point on the curve $\vec{u}(\varphi) = \vec{c}(s)$ that is \textit{not} singular), hence (later dropping the $\tilde{•}$ )
\begin{equation}\label{eq5}
cos \phi = \tilde{r}' \left( =\dfrac{d\tilde{r}}{ds} \right),
\end{equation}
(following by differentiating $\tilde{r}^2 = \vec{c}\,^2$ with respect to $s$ and using $|\vec{c}\,'(s)| = 1; r(\varphi) = \tilde{r}(s(\varphi))$) finding that
%%%%%%%%%%%%%%%%%%%%%%%%%%%%%%%%%%%%%%%%%%%%%%%%%%%%%%%%%%%%%%%%%%%%%%%%%%%%%
%PAGE 4 PAGE 4 PAGE 4 PAGE 4 PAGE 4 PAGE 4 PAGE 4 PAGE 4 PAGE 4 PAGE 4 PAGE 4
%%%%%%%%%%%%%%%%%%%%%%%%%%%%%%%%%%%%%%%%%%%%%%%%%%%%%%%%%%%%%%%%%%%%%%%%%%%%%
\begin{equation}\label{eq6}
\omega^2 r^2 = \gamma_0^2 + \dfrac{\omega^2}{1-\gamma_0^2}(s-s_0)^2\; (= \gamma_0^2 + \beta \omega^2 s^2);
\end{equation}
$s_0$ can be (and is) put to zero by starting at a point of minimal distance $r_{\text{Min}}$ from the origin when $\gamma_0^2 \in (0,1)$, and at a point of maximal distance $r_{\text{Max}}$ when $\gamma_0^2 > 1$ (and, following the curve in both positive and negative $s$ direction, reaching a singularity at $|s|_{\text{Max}} = \hat{s}$). Instead of calculating $\theta(s)$ from $(\vec{c}\,'^2 = 1)$
\begin{equation}\label{eq7}
r^2 \theta'^2 = (1 - r'^2)
\end{equation}
i.e. from
\begin{equation}\label{eq8}
\pm \theta' = \dfrac{\sqrt{1-r'^2}}{r} = |\gamma_0|\dfrac{\sqrt{1-\beta^2\omega^2 s^2}}{(\gamma_0^2+\beta\omega^2 s^2)}
\end{equation}
use (\ref{eq7}), to calculate the curvature $\kappa$ from $\vec{c}\,' = r'
\begin{pmatrix}
c \\ s
\end{pmatrix}
+ r\theta'
\begin{pmatrix}
-s \\ c
\end{pmatrix}
=: \vec{e}_1
$,
$\vec{e}\,'_1 = (r'' - r\theta'^2)
\begin{pmatrix}
c \\ s
\end{pmatrix} + (r\theta'' + 2r'\theta')
\begin{pmatrix}
-s \\ c
\end{pmatrix}
\stackrel{!}{=}
\kappa \vec{e}_2
$, implying (as $\vec{e}_2 = (-r\theta')
\begin{pmatrix}
c \\ s
\end{pmatrix}
+ r'
\begin{pmatrix}
-s \\ c
\end{pmatrix}
$
gives $det(\vec{e}_1,\vec{e}_2) = +1$)
\begin{equation}\label{eq9}
\kappa = \dfrac{(r\theta'^2 -r'')}{r\theta'} = \mp \left( \dfrac{r''}{\sqrt{1-r'^2}} - \dfrac{\sqrt{1-r'^2}}{r}\right)
= \dfrac{\kappa_0}{\sqrt{1-\mu^2s^2}}
\end{equation}
where/with $\mu^2 = \omega^2\beta^2,\; \kappa_0 = \mp |\gamma_0| |\omega| \beta (= -\gamma_0 \omega \beta,$ s.b.)
%%%%%%%%%%%%%%%%%%%%%%%%%%%%%%%%%%%%%%%%%%%%%%%%%%%%%%%%%%%%%%%%%%%%%%%%%%%%%
%PAGE 5 PAGE 5 PAGE 5 PAGE 5 PAGE 5 PAGE 5 PAGE 5 PAGE 5 PAGE 5 PAGE 5 PAGE 5
%%%%%%%%%%%%%%%%%%%%%%%%%%%%%%%%%%%%%%%%%%%%%%%%%%%%%%%%%%%%%%%%%%%%%%%%%%%%%
using (\ref{eq6}), resp. ($\beta = \frac{1}{1-\gamma_0^2}, \omega >0$)
\begin{equation}\label{eq10}
r'' = \dfrac{\omega \beta \gamma_0^2}{(\omega r)^3},\; r' = \dfrac{\beta |\omega| s}{\sqrt{\gamma_0^2 + \beta(\omega^2 s^2)}} = \beta \dfrac{|\omega| s}{r}\gtrless 0
\begin{pmatrix}
\gamma_0^2 < 1 \\
\gamma_0^2 > 1
\end{pmatrix}
\end{equation}
Concerning signs, it is convenient to choose $+$ in (\ref{eq8}), i.e. letting positive $s$ correspond to moving in the mathematically positive direction on the curve (i.e. against the clock). For $\gamma_0^2 > 1\; (\beta <0)$ one then has $\kappa_0(s) > 0$ for $s \in (-\hat{s}, +\hat{s}) =: I$ and  $\kappa_0(s) < 0 \; (s \in I)$ for $\gamma_0^2 < 1 \; (\beta > 0)$. Also, as (always, in any parametrization) $\vec{u}\times \vec{u}' = r^2 \theta' = r |\vec{u}'| sin \phi$, $sin \phi \geqslant 0$ by definition $(\theta' \geqslant 0)$, hence $\gamma_0$ and $\omega$ (cp.(\ref{eq4})) having the same sign in this convention, so that the absolute value signs in (\ref{eq9}) can be dropped, i.e.
\begin{equation}\label{eq11}
\kappa_0 = -\gamma_0 \omega \beta.
\end{equation}
Both (\ref{eq8}) and (\ref{eq9}) show that the curve becomes singular when $\omega s \rightarrow \pm (1- \gamma_0^2)$ i.e. at $\omega^2 r^2 = 1$ (resp. $\phi = \arcangle (\vec{u}, \vec{u}') = 0, \theta ' = 0$) when $\kappa$ becomes infinite.
%%%%%%%%%%%%%%%%%%%%%%%%%%%%%%%%%%%%%%%%%%%%%%%%%%%%%%%%%%%%%%%%%%%%%%%%%%%%%
%PAGE 6 PAGE 6 PAGE 6 PAGE 6 PAGE 6 PAGE 6 PAGE 6 PAGE 6 PAGE 6 PAGE 6 PAGE 6
%%%%%%%%%%%%%%%%%%%%%%%%%%%%%%%%%%%%%%%%%%%%%%%%%%%%%%%%%%%%%%%%%%%%%%%%%%%%%
Having determined the curvature $\kappa$ of the curve, one may notice that the local radius of curvature, $\rho := \frac{1}{x}$, and the arclength parameter $s$ lie on an ellipse,
\begin{equation}\label{eq12}
\dfrac{s^2}{p^2} + \dfrac{\rho^2}{q^2} = 1
\end{equation}
with $q^2 = \frac{1}{\kappa_0^2} = \frac{(1-\gamma_0^2)^2}{\gamma_0^2\omega^2}$ and $p^2 =  \frac{(1-\gamma_0^2)^2}{\omega^2}$, i.e. the ratio of the defining half-axis being $|\gamma_0|$. This is the second chance to consult classical geometry, resp. to find [`Epicycloid', mathworld.wolfram.com] that rolling a circle of radius $q$ around a circle of radius $p$ one will get a curve (an `epicycloid') with exactly that curvature $\kappa$ (cp. (\ref{eq9})), resp. satisfying (\ref{eq12}). While, given this observation, it should be even easier to arrive at these curves by an elementary geometric argument directly from (\ref{eq2}),  let us once again choose to go another route, namely using the standard integration of the Frenet-frame ODE to deduce the planar curve from its curvature.\\\\
%%%%%%%%%%%%%%%%%%%%%%%%%%%%%%%%%%%%%%%%%%%%%%%%%%%%%%%%%%%%%%%%%%%%%%%%%%%%%
%PAGE 7 PAGE 7 PAGE 7 PAGE 7 PAGE 7 PAGE 7 PAGE 7 PAGE 7 PAGE 7 PAGE 7 PAGE 7
%%%%%%%%%%%%%%%%%%%%%%%%%%%%%%%%%%%%%%%%%%%%%%%%%%%%%%%%%%%%%%%%%%%%%%%%%%%%%
Calculating
\begin{equation}\label{eq13}
\sigma(s) := \int_0^s\kappa(v)dv = \dfrac{\kappa_0}{\mu}\int_0^{\mu s}\dfrac{dw}{\sqrt{1-w^2}} = \dfrac{\kappa_0}{\mu}arcsin(\mu s)
\end{equation}
one has (with $\mu t = sin(\mu u),\, \mu s = sin (\mu w)$)
\begin{equation}\label{eq14}
\begin{split}
\vec{u}(\varphi) &= \vec{c}(s)\\
  & = \vec{c}(0) + \int_0^s
\begin{pmatrix}
cos(\sigma(t)+\sigma_0) \\ sin(\sigma(t)+\sigma_0)
\end{pmatrix} dt \\
  & = \int_0^{\frac{1}{\mu}arcsin(\mu s)}cos(\mu u)
\begin{pmatrix}
cos(\kappa_0 u+\sigma_0) \\ sin(\kappa_0 u+\sigma_0)
\end{pmatrix} du + \vec{c}(0) \\
 & = \dfrac{1}{\mu^2-\kappa_0^2}
 \begin{pmatrix}
 \mu sin(\mu u)cos(\kappa_0 u + \sigma_0) - \kappa_0cos(\mu u)sin(\kappa_0 u +\sigma_0)\\
 \mu sin(\mu u)sin(\kappa_0 u + \sigma_0) + \kappa_0cos(\mu u)cos(\kappa_0 u +\sigma_0)
 \end{pmatrix}\bigg\vert_0^w
 + \vec{c}(0)\\
 & = \dfrac{(1-\gamma_0^2)}{\omega^2}R(\sigma_0)R(\kappa_0 u)
 \begin{pmatrix}
 \mu sin(\mu u) \\ \kappa_0 cos(\mu u)
 \end{pmatrix} \bigg\vert_0^w
 + \vec{c}(0)\\
 & = -\dfrac{1}{\omega} R(\sigma_0)R(+\gamma_0 v) \underbrace{
 \begin{pmatrix}
 sin (v) \\ +\gamma_0 cos(v)
 \end{pmatrix}
 }_ {=: \vec{w}\omega}
+ \underbrace{\vec{c}(0) - \dfrac{1}{\omega}R(\sigma_0)
\begin{pmatrix}
0 \\ -\gamma_0
\end{pmatrix}
}_{ =: \vec{d}}
\end{split}
\end{equation}
where $v = -\mu w = -arcsin(\mu s)$,
hence (if $\vec{d} = \vec{0}$)
\begin{equation}\label{eq15}
x = \begin{pmatrix}
t \\ -R(\omega t + \gamma_0 v + \sigma_0)\vec{v}(v)
\end{pmatrix}
= \begin{pmatrix}
au - bv -a\sigma_0 \\ -R(u)\vec{w}(v)
\end{pmatrix}
= x(u,v)
\end{equation}
where $u = \omega t + \gamma_0 v + \sigma_0, \; a =\frac{1}{\omega}, \; b = \frac{\gamma_0}{\omega} = \gamma_0 a$.
%%%%%%%%%%%%%%%%%%%%%%%%%%%%%%%%%%%%%%%%%%%%%%%%%%%%%%%%%%%%%%%%%%%%%%%%%%%%%
%PAGE 8 PAGE 8 PAGE 8 PAGE 8 PAGE 8 PAGE 8 PAGE 8 PAGE 8 PAGE 8 PAGE 8 PAGE 8
%%%%%%%%%%%%%%%%%%%%%%%%%%%%%%%%%%%%%%%%%%%%%%%%%%%%%%%%%%%%%%%%%%%%%%%%%%%%%
Again a comment about signs\footnote{one should think it to be overly pedantic to worry about sign conventions in such an elementary context, but it not only turned out to be a nuisance to not be able to rely on the signs in some of the equations, but also, curiously and unexpectedly often, led to confusion despite of the apparent simplicity of the matter.}: while the space-inversion $(-\frac{1}{\omega}\ldots)$ is harmless (as it is orientation-preserving) $v$ and $s$, if taking $\mu = \omega\beta$ (no absolute signs, so $\mu$ may be negative), have  (for $\omega > 0$) the same sign for $\gamma_0^2 > 1$ while opposite signs for $\gamma_0^2 < 1$. For small $v$,
\begin{equation}\label{eq16}
R(\gamma_0 v)\begin{pmatrix}
sin(v) \\ \gamma_0 cos(v)
\end{pmatrix}
= \begin{pmatrix}
(1-\gamma_0^2)v + o(v)^3 \\
\gamma_0[1+(1-\gamma_0^2)\frac{v^2}{2}] + o(v^4)
\end{pmatrix}
= -\omega\vec{u}(v)
\end{equation}
\begin{figure}[h]
\centering
\includegraphics[width=10.50cm]{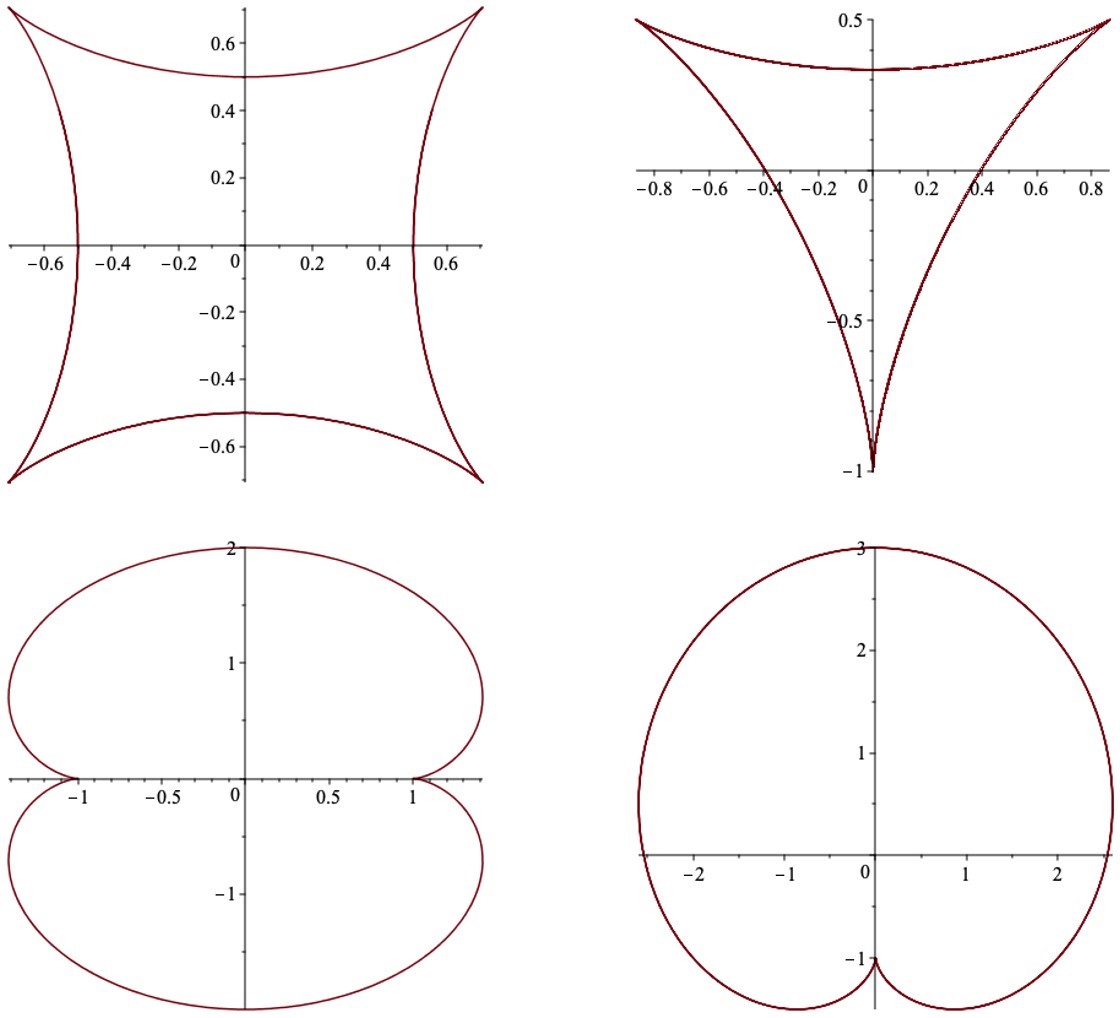}
\end{figure}\\
Looking at the pictures\footnote{many thanks to J.Eggers and M.Hynek, as well as G.Linardopoulos} for $\gamma_0 = \frac{1}{2}, \frac{1}{3},2, 3$ (always starting on the positive $y$-axis) $s>0$ always (consistent with the chosen conventions, $\frac{d\theta}{ds} > 0$) moves to the left, with $\kappa(0) = \frac{-\omega\gamma_0}{1-\gamma_0^2}$ (cp. (\ref{eq9})) being negative for $\gamma_0^2 < 1$ and positive for $\gamma_0^2 >1$ (note that $\frac{\vec{u}'(v)\times\vec{u}''(v)}{|\vec{u}'(v)|^3}\vert_{v=0}$ gives $\frac{\omega\gamma_0}{|1-\gamma_0^2|}$) which seems to give the right result only for $\gamma_0^2>1$; for $\gamma_0^2<1$, however, the orientation reversing transformation $v\leftrightarrow s$ accounts for an extra-sign)
%%%%%%%%%%%%%%%%%%%%%%%%%%%%%%%%%%%%%%%%%%%%%%%%%%%%%%%%%%%%%%%%%%%%%%%%%%%%%
%PAGE 9 PAGE 9 PAGE 9 PAGE 9 PAGE 9 PAGE 9 PAGE 9 PAGE 9 PAGE 9 PAGE 9 PAGE 9
%%%%%%%%%%%%%%%%%%%%%%%%%%%%%%%%%%%%%%%%%%%%%%%%%%%%%%%%%%%%%%%%%%%%%%%%%%%%%
One of the pleasant surprises of having gone through the tedious derivation of (\ref{eq15}), instead of having employed elementary geometry right at the beginning (cp.(\ref{eq2})), is that one automatically arrives at the conformal (!) parametrization ($a = \frac{1}{\omega}$, $b = \frac{\gamma_0}{\omega}$)
\begin{equation}\label{eq17}
\begin{split}
x(u,v) & = - \begin{pmatrix}
-au + bv \\ a cos(u) sin(v)- b sin(u)cos(v)\\ a sin(u) sin(v) + b cos(u)cos(v)
\end{pmatrix} \\
 & = -a \begin{pmatrix}
-u + \gamma_0 v \\ cos(u)sin(v) - \gamma_0 sin(u)cos(v) \\ sin(u)sin(v)+ \gamma_0 cos(u)cos(v)
\end{pmatrix}  \\
 & = \begin{pmatrix}
 au-bv \\ -R(u)\underbrace{\begin{pmatrix}
 a sin(v) \\ b cos(v)
 \end{pmatrix}}_{=: \vec{w}(v)}
 \end{pmatrix}\\
 & = \dfrac{(a-b)}{2}\begin{pmatrix}
 u+v \\ -sin(u+v) \\cos(u+v)
 \end{pmatrix} + \dfrac{a+b}{2} \begin{pmatrix}
 u-v \\ sin(u-v) \\-cos(u-v)
 \end{pmatrix},
\end{split}
\end{equation}
the sum of two null-curves, just as ($-R(\sigma_0)$ of) the planar curve that is being rotated,
\begin{equation}\label{eq18}
\begin{split}
\vec{r}(v) & = R(\gamma_0v)\begin{pmatrix}
a sin(v) \\ b cos(v)
\end{pmatrix}\\
 & =\dfrac{a+b}{2}\begin{pmatrix}
 -sin (\frac{b-a}{a}v) \\ cos (\frac{b-a}{a}v)
 \end{pmatrix} + \dfrac{b-a}{2}\begin{pmatrix}
 -sin (\frac{b+a}{a}v) \\ cos (\frac{b+a}{a}v)
 \end{pmatrix}\\
 & = \dfrac{1}{\omega}(\gamma_0^2-1)\left\lbrace \dfrac{1}{2(\gamma_0-1)}\begin{pmatrix}
 -sin((\gamma_0 -1)v) \\ cos((\gamma_0 -1)v)
 \end{pmatrix} +
 \dfrac{1}{2(\gamma_0+1)}\begin{pmatrix}
 -sin((\gamma_0+1)v) \\ cos((\gamma_0+1)v)
 \end{pmatrix}\right\rbrace
\end{split}
\end{equation}
%%%%%%%%%%%%%%%%%%%%%%%%%%%%%%%%%%%%%%%%%%%%%%%%%%%%%%%%%%%%%%%%%%%%%%%%%%%%%
%PAGE 10 PAGE 10 PAGE 10 PAGE 10 PAGE 10 PAGE 10 PAGE 10 PAGE 10 PAGE 10 PAGE 10
%%%%%%%%%%%%%%%%%%%%%%%%%%%%%%%%%%%%%%%%%%%%%%%%%%%%%%%%%%%%%%%%%%%%%%%%%%%%%
At this stage it may be useful\footnote{the reader interested only in the result(concerning what the curves described by (\ref{eq18}) are) may skip the following, somewhat lengthy detour, and go directly to eq.(30).} to have a look at solutions that were given in \cite{1} (eq.(52) ; note that in (53) the following typo should be corrected: the rhs is the square(!) of the radius $r$, not $r$),
\begin{equation}\label{eq19}
\vec{x}(t, \tilde{\varphi}) = \dfrac{1}{2m} \begin{pmatrix}
cos(m(\tilde{\varphi}+t)) \\ sin (m(\tilde{\varphi}+t))
\end{pmatrix} + \dfrac{1}{2n} \begin{pmatrix}
cos(n(\tilde{\varphi}-t)) \\ sin (n(\tilde{\varphi}-t))
\end{pmatrix}
\end{equation}
(to be compared with the rhs of (\ref{eq17}), which has $au-bv = \frac{a-b}{2}(u+v)- \frac{a+b}{2}(v-u)$ as first component, just as $\frac{1}{2m}(m(\tilde{\varphi}+t)) - \frac{1}{2n}(n(\tilde{\varphi}-t)) =t$), corresponding to choosing $f = \frac{m}{2}(\tilde{\varphi}+t), \, g = \frac{n}{2}(\tilde{\varphi}-t)$ in (50). Calculating the curvature of the time-dependent planar curve (\ref{eq19}) gives (cp.(51))
\begin{equation}\label{eq20}
\begin{split}
\tilde{\kappa}(\vec{\varphi},t)& = \dfrac{m+n}{2cos(f-g)} \\
 & = \dfrac{m+n}{2cos(\frac{m-n}{2}\tilde{\varphi}+\frac{m+n}{2}t)}\\
 & = \dfrac{m+n}{2cos(\frac{m-n}{2}\varphi)}\\
 & =: \kappa(\varphi)
\end{split}
\end{equation}
upon reparametrizing (cp.(53)), $\varphi :=  \tilde{\varphi} + (\frac{m+n}{m-n})t$ (in any case, as $\varphi$ and $\tilde{\varphi}$ coincide at $t = 0$, $\kappa(\varphi)$ is the curvature of the curve $\vec{u}(\varphi) := \vec{x}(t=0, \tilde{\varphi}= \varphi)$ that is being rotated with constant angular velocity (cp.II.74 \cite{7} $\varphi \leftrightarrow \tilde{\varphi}$) to give (\ref{eq19})).
%%%%%%%%%%%%%%%%%%%%%%%%%%%%%%%%%%%%%%%%%%%%%%%%%%%%%%%%%%%%%%%%%%%%%%%%%%%%%
%PAGE 11 PAGE 11 PAGE 11 PAGE 11 PAGE 11 PAGE 11 PAGE 11 PAGE 11 PAGE 11 PAGE 11
%%%%%%%%%%%%%%%%%%%%%%%%%%%%%%%%%%%%%%%%%%%%%%%%%%%%%%%%%%%%%%%%%%%%%%%%%%%%%
As \begin{equation}\label{eq21}
\begin{split}
(\vec{u}\,'(\varphi))^2 & = \left( \dfrac{1}{2}\begin{pmatrix}
cos(m\varphi)\\sin(m\varphi)
\end{pmatrix} + (\dfrac{1}{2}\begin{pmatrix}
cos(n\varphi)\\sin(n\varphi)
\end{pmatrix}\right)^2 \\
 & = \dfrac{1}{2}(1 + (c_m c_n + s_m s_n)) \\
 & = cos^2\left( \dfrac{m-n}{2}\varphi\right) ,
\end{split}
\end{equation}
\begin{equation}\label{eq22}
s(\varphi) = \int_0^{\varphi}|cos \left( \dfrac{m-n}{2}u\right)|du
= \dfrac{2}{m-n}sin\left( \dfrac{m-n}{2}\varphi\right)
\end{equation}
for small enough $\varphi$. Hence
\begin{equation}\label{eq23}
\kappa(s) = \dfrac{m+n}{2\sqrt{1-\frac{(m-n)^2}{4}s^2}},
\end{equation}
i.e. $\kappa_0 = \frac{m+n}{2}$, $\mu^2 = \frac{(m-n)^2}{4}$, hence
\begin{equation}\label{eq24}
\gamma_0^2 = \dfrac{\kappa_0^2}{\mu^2} = \dfrac{(m+n)^2}{(m-n)^2}.
\end{equation}
The point of going through this explicit, simple, calculation is to see in this way that all rational values for $\gamma_0$ correspond to simple, closed curves of the form (\ref{eq19}), that $\gamma_0^2$ being bigger or smaller $1$ depends on whether $m$ and $n$ have the \textit{same}, or opposite signs, and that $m\rightarrow \lambda m$, $n\rightarrow \lambda n$ does not change $\gamma_0$ (consistent with \cite{1}, where the reason for choosing $m$ and $n$ to have no common divisor $\neq 1$ had been to ensure that $\varphi \in [0,2\pi]$ is the minimal periodicity interval).
%%%%%%%%%%%%%%%%%%%%%%%%%%%%%%%%%%%%%%%%%%%%%%%%%%%%%%%%%%%%%%%%%%%%%%%%%%%%%
%PAGE 12 PAGE 12 PAGE 12 PAGE 12 PAGE 12 PAGE 12 PAGE 12 PAGE 12 PAGE 12 PAGE 12
%%%%%%%%%%%%%%%%%%%%%%%%%%%%%%%%%%%%%%%%%%%%%%%%%%%%%%%%%%%%%%%%%%%%%%%%%%%%%
Letting $\gamma_0 = \frac{m+n}{m-n}$ in (\ref{eq18}), one gets (with $\frac{v}{m-n} = 2\varphi$)
\begin{equation}\label{eq25}
\omega \vec{r} = \dfrac{mn}{(m-n)}\left( \dfrac{1}{n}\begin{pmatrix}
-s_n \\ c_n
\end{pmatrix} + \dfrac{1}{m}\begin{pmatrix}
-s_m \\ c_m
\end{pmatrix} \right)
\end{equation}
where $s_k = sin(k\varphi), c_k = cos(k\varphi)$.\\
Now for the interpretation in terms of rolling circles around (or inside) circles. As derived e.g. in [Wikipedia] (cp.also II.76); note in the second line the missing letter $R$ in front of $\frac{b}{a}\psi$, and that the big closing bracket should not be at the end, but after $\vec{b} = \big(\begin{smallmatrix}
b \\ 0
\end{smallmatrix}\big)$)
parametric descriptions are given by
\begin{equation}\label{eq26}
\vec{x}(\psi) = \rho(q+1)\left( \begin{pmatrix}
cos(\psi) \\ sin (\psi)
\end{pmatrix} - \dfrac{1}{q+1} \begin{pmatrix}
cos((q+1)\psi)\\ sin((q+1)\psi)
\end{pmatrix} \right)
\end{equation}
for epicycloids (a circle of radius $\rho$ rolling on the outside of a circle of radius $\rho q > \rho$), and
\begin{equation}\label{eq27}
\vec{x}(\psi) = \rho(q-1)\left( \begin{pmatrix}
cos(\psi) \\ sin (\psi)
\end{pmatrix} + \dfrac{1}{q-1} \begin{pmatrix}
cos((q-1)\psi)\\ -sin((q-1)\psi)
\end{pmatrix} \right)
\end{equation}
for hypocycloids (a circle of radius $\rho$ rolling on the inside of a circle of radius $\rho q > \rho$). Comparing (\ref{eq26}) with (\ref{eq25}) ($m > n > 0$): put $-n\varphi = \psi$, $-m\varphi = \frac{m}{n}\psi$ $\Rightarrow q+1 = \frac{m}{n}$, i.e. the fixed circle is $q= \frac{m-n}{n}$ times bigger; in the 2 examples : $\gamma_0 = 3,\,m=2,\,n=1,\,q=1$ (same size) $\Rightarrow$ cardioid with $\rho_3 = \frac{n}{m-n} = 1$, $\rho_2 = \frac{1}{2}$  $\gamma_0 = 2,\,m=3,\,n=1,\,q=2$ (nephroid),
%%%%%%%%%%%%%%%%%%%%%%%%%%%%%%%%%%%%%%%%%%%%%%%%%%%%%%%%%%%%%%%%%%%%%%%%%%%%%
%PAGE 13 PAGE 13 PAGE 13 PAGE 13 PAGE 13 PAGE 13 PAGE 13 PAGE 13 PAGE 13 PAGE 13
%%%%%%%%%%%%%%%%%%%%%%%%%%%%%%%%%%%%%%%%%%%%%%%%%%%%%%%%%%%%%%%%%%%%%%%%%%%%%
consistent with the picture ($d = \gamma_0$) obtained from $R(\gamma_0v)\big(\begin{smallmatrix}
sin(v)\\ \gamma_0cos(v)
\end{smallmatrix}\big)$. Note that in order for the identification to work one has to interchange $x$ and $y$ coordinates (a reflexion around the diagonal, which does not change the nature of the shape), as well as argue why the relative $-$sign in (\ref{eq26}) between the 2 vectors should not matter when comparing with (\ref{eq25}): as shown in II.78 of \cite{7} the $-$sign can be transformed into a plus sign by simultaneously shifting the angle by $\frac{\pi}{m-n}$ and rotating by $\frac{m}{m-n}\pi$. For $\gamma_0^2 <1$
, $\gamma_0 = \frac{m+n}{m-n}$ with $n = -n' <0$ (and $m \geqslant n'$) (\ref{eq25}) gives
\begin{equation}\label{eq28}
\begin{split}
\omega \vec{r}(v) & = \dfrac{mn'}{m+n'}\left( \dfrac{1}{n'}\begin{pmatrix}
s_{n'} \\ c_{n'}
\end{pmatrix} - \dfrac{1}{m}\begin{pmatrix}
-s_{m} \\ +c_{m}
\end{pmatrix} \right) (v) \\
 & = \dfrac{m}{m+n'}\left(\begin{pmatrix}
s \\ c
\end{pmatrix} - \dfrac{n'}{m}\begin{pmatrix}
-s_{\frac{m}{n'}} \\ +c_{\frac{m}{n'}}
\end{pmatrix} \right) (v' = n'v)
\end{split}
\end{equation}
i.e. (again, the relative $-$sign transformable into a $+$sign s.a.) $q-1 = \frac{m}{n'}$, i.e.
\begin{equation}\label{eq29}
q = \dfrac{m+n'}{n'},\; \rho = \dfrac{n'}{m+n'}.
\end{equation}
%%%%%%%%%%%%%%%%%%%%%%%%%%%%%%%%%%%%%%%%%%%%%%%%%%%%%%%%%%%%%%%%%%%%%%%%%%%%%
%PAGE 14 PAGE 14 PAGE 14 PAGE 14 PAGE 14 PAGE 14 PAGE 14 PAGE 14 PAGE 14 PAGE 14
%%%%%%%%%%%%%%%%%%%%%%%%%%%%%%%%%%%%%%%%%%%%%%%%%%%%%%%%%%%%%%%%%%%%%%%%%%%%%
In the examples : $\gamma_0 = \frac{1}{2} \, (m = 3, n' =1)$ gives $q=4, \rho =\frac{1}{4}$, while $\gamma = \frac{1}{3} \, (m=2, n'=1)$ gives $q = 3, \rho = \frac{1}{3}$ (consistent with the pictures); the bigger circle always having unit radius (for the epicycloids : $q+1 = \frac{m}{n}$ , $q = \frac{m-n}{n}$, $\rho = \frac{n}{m-n}$ being the radius of the circle rolling around the unit circle; note that $\rho$ can be both smaller, equal, or bigger than $1$, while in the case of hypocycloids, $\gamma_0^2 < 1, \, \rho$ obviously can be at most $\frac{1}{2}$ and in that limiting case, $\gamma_0 = 0$, giving a straight line - which is the first, 2 cusped,  `Tusi-couple' hypocycloids, named after the Persian astronomer and mathematician Nasir al-Din al-Tusi, who discussed in 1247 the linear motion arising from a circle rolling inside a circle of twice its radius). The case of negative $\gamma_0$ does not need to be separately discussed, as
\begin{equation}\label{eq30}
\vec{r}_{-\gamma_0}(v) = -\vec{r}_{\gamma_0}(-v).
\end{equation}
%%%%%%%%%%%%%%%%%%%%%%%%%%%%%%%%%%%%%%%%%%%%%%%%%%%%%%%%%%%%%%%%%%%%%%%%%%%%%
%PAGE 15 PAGE 15 PAGE 15 PAGE 15 PAGE 15 PAGE 15 PAGE 15 PAGE 15 PAGE 15 PAGE 15
%%%%%%%%%%%%%%%%%%%%%%%%%%%%%%%%%%%%%%%%%%%%%%%%%%%%%%%%%%%%%%%%%%%%%%%%%%%%%
Summarizing the result (which of course can be seen much more simply, directly from (\ref{eq18})): Each member of the one-parameter class of curves ($-R(-\sigma_0)$ times)
\begin{equation}\label{eq31}
\omega\vec{r}(v) = R(\gamma_0 v)\begin{pmatrix}
sin(v) \\ \gamma_0 cos(v)
\end{pmatrix}  = \vec{v}_{\gamma_0}(v)
\end{equation}
(because of (\ref{eq30}) assume $\gamma_0 \geqslant 0$ from now on), that when divided by $\omega$ and rotated with constant angular velocity $\omega$ gives rise to a time-like (apart from the singularities) `minimal' surface in $\mathbb R^{1,2}$, is obtained by rolling a circle of radius $\rho = \frac{1}{2}|1-\gamma_0|$ on the outside (if $\gamma_0 > 1$) or the inside (if $\gamma_0 < 1$) of a unit circle. For rational $\gamma_0 \in [-3,+3],\, \gamma_0^2 \neq 1$, the range of $v$ can be chosen such that $\vec{v}_{\gamma_0}(v)$ describes a closed curve without self-intersections (the companion with the $m,n$ discussion is simple, as e.g. $\gamma_0 = \frac{p}{k} = \frac{m+n}{m-n}$ for $p > k > 0, \, m > n > 0$ implies $n = p-k,\, m = p+k$ if $n$ and $m$, as well as $p$ and $k$ have $1$ as their biggest common factor). \\
%%%%%%%%%%%%%%%%%%%%%%%%%%%%%%%%%%%%%%%%%%%%%%%%%%%%%%%%%%%%%%%%%%%%%%%%%%%%%
%PAGE 16 PAGE 16 PAGE 16 PAGE 16 PAGE 16 PAGE 16 PAGE 16 PAGE 16 PAGE 16 PAGE 16
%%%%%%%%%%%%%%%%%%%%%%%%%%%%%%%%%%%%%%%%%%%%%%%%%%%%%%%%%%%%%%%%%%%%%%%%%%%%%
Calculating the metric in the parametrization
\begin{equation}\label{eq32}
x(t, v) = \begin{pmatrix}
t \\ -\dfrac{1}{\omega}(\sigma_0)R(\omega t)R(\gamma_0 v)\begin{pmatrix}
sin(v) \\ \gamma_0 cos(v)
\end{pmatrix}
\end{pmatrix} = \begin{pmatrix}
t \\ \vec{x}(t,v)
\end{pmatrix}
\end{equation}
one gets
\begin{equation}\label{eq33}
(G_{\alpha \beta}) = \begin{pmatrix}
\dot{x}^2 & \dot{x}x' \\
\cdot & x'^2
\end{pmatrix} = \dfrac{(1-\gamma_0^2)}{\omega} cos^2(v) \begin{pmatrix}
\omega & \gamma_0 \\ \gamma_0 & \frac{(\gamma_0^2-1)}{\omega}
\end{pmatrix},
\end{equation}
which \textit{is} conformally constant, but (at first alarmingly) has $\dot{x}^2 = 1-\dot{\vec{x}}^2 < 0$ for $\gamma_0^2 > 1$, (meaning that points of fixed $v$ move with a velocity $> 1$)\footnote{The resolution of this puzzle rests in the relation between $\tilde{\varphi}$ and $\varphi$ (cp.(\ref{eq20})); (\ref{eq19}) does satisfy $1-\dot{\vec{x}}^2 = \vec{x}'^2 > 0$.}; the constant matrix on the rhs of (\ref{eq33}) however does have determinant $-1$ (for all $\gamma_0$), meaning (as the sign of the prefactor is irrelevant for the determinant) that the surface  in $\mathbb R^{1,2}$ \textit{is} time-like,(for all $\gamma_0^2 \neq 1$ and $cos(v) \neq 0$). In $u, v$ coordinates,
\begin{equation}\label{eq34}
(\tilde{G}_{\alpha \beta}) = \begin{pmatrix}
1 & 0 \\ 0 &-1
\end{pmatrix} (cos^2 (v))\underbrace{(a^2 - b^2)}_{= \frac{1}{\omega^2}(1-\gamma_0^2)},
\end{equation}
verifying the conformality of the $(u,v)$ parametrization (\ref{eq17}).
%%%%%%%%%%%%%%%%%%%%%%%%%%%%%%%%%%%%%%%%%%%%%%%%%%%%%%%%%%%%%%%%%%%%%%%%%%%%%
%PAGE 17 PAGE 17 PAGE 17 PAGE 17 PAGE 17 PAGE 17 PAGE 17 PAGE 17 PAGE 17 PAGE 17
%%%%%%%%%%%%%%%%%%%%%%%%%%%%%%%%%%%%%%%%%%%%%%%%%%%%%%%%%%%%%%%%%%%%%%%%%%%%%
As a consistency-check one can verify that
\begin{equation}\label{eq35}
J^T\begin{pmatrix}
1 & 0 \\ 0 & -1
\end{pmatrix}J  = \omega \begin{pmatrix}
\omega & \gamma_0 \\ \gamma_0 & \frac{\gamma_0^2-1}{\omega}
\end{pmatrix} \:
\text{for} \: J  = \begin{pmatrix}
\omega & \gamma_0 \\ 0 & 1
\end{pmatrix} = \dfrac{\partial(u,v)}{\partial(t,v)}.
\end{equation}
Also, just as (\ref{eq17}) obviously satisfies
\begin{equation}\label{eq36}
(\partial_u^2 - \partial_v^2)x(u,v) = 0,
\end{equation}
one can easily check that the curve $\vec{v}$ that is being rotated satisfied the \textit{linear} ODE
\begin{equation}\label{eq37}
\vec{v}'' - 2 \gamma_0A\vec{v}' + (1-\gamma_0^2)\vec{v} = \vec{0},
\end{equation}
which is interesting in its own right as the original shape equation, and minimality conditions, were \textit{non}-linear. One should also note the lucky `coincidence' that the constant vector $\vec{d} := \vec{c}(0) - \frac{1}{\omega}R(\sigma_0)\big(\begin{smallmatrix}
o \\ -\gamma_0
\end{smallmatrix}\big) $
is identically zero (given the curvature, (\ref{eq9}), the curve is given only up to rotations -in the case at hand harmless- and translations; due to prefactor $R(\omega t)$ it is \textit{not} true that (\ref{eq14}) would satisfy the shape equation irrespective of what $\vec{d}$ is; with less `luck' there could have been additional work needed to find out the correct $\vec{d}$; due to (\ref{eq36})/(\ref{eq37}) this is fortunately not necessary).\\
%%%%%%%%%%%%%%%%%%%%%%%%%%%%%%%%%%%%%%%%%%%%%%%%%%%%%%%%%%%%%%%%%%%%%%%%%%%%%
%PAGE 18 PAGE 18 PAGE 18 PAGE 18 PAGE 18 PAGE 18 PAGE 18 PAGE 18 PAGE 18 PAGE 18
%%%%%%%%%%%%%%%%%%%%%%%%%%%%%%%%%%%%%%%%%%%%%%%%%%%%%%%%%%%%%%%%%%%%%%%%%%%%%
Yet another solution: we solved the shape-equation by deriving the curvature in arc-length-parametrization, and  then integrating the Frenet equations, finding as a general solution of
\begin{equation}\label{eq38}
\gamma_0^2\vec{v}'^2(1-\vec{v}\,^2) = (1-\gamma_0^2)(\vec{v} \times \vec{v}')^2
\end{equation}
\begin{equation}\label{eq39}
\begin{split}
\vec{v}(v) & = R(\gamma_0 v)\begin{pmatrix}
sin(v) \\ \gamma_0 cos(v)
\end{pmatrix}\\
 & = \dfrac{\gamma_0 +1}{2}\begin{pmatrix}
 -s_- \\c_-
 \end{pmatrix} + \dfrac{\gamma_0 -1}{2}\begin{pmatrix}
 -s_+ \\c_+
 \end{pmatrix},
\end{split}
\end{equation}
$s_{\pm} = sin(\gamma_0 \pm 1)v, \, c_{\pm} = cos(\gamma_0 \pm 1)v$ (and up to rotating or reflecting $\vec{v}$ and $v\rightarrow v_0$, not having been particularly careful about that constant, resp. originally having defined $v= -arcsin(\mu s)$, with $s=0$ furthest away from the next singularity); note that (\ref{eq38}), as it must be in order to be a `geometric' equation (i.e. for the \textit{shape} of the curve), is reparametrization invariant. Curiously the general solution of (\ref{eq38}) turned out, in a particular parametrization $v$, s.a., to satisfy a \textit{linear} ODE, (\ref{eq37}), ($\vec{v}'' - 2 \gamma_0A\vec{v}' + (1-\gamma_0^2)\vec{v} = \vec{0}$).\\
%%%%%%%%%%%%%%%%%%%%%%%%%%%%%%%%%%%%%%%%%%%%%%%%%%%%%%%%%%%%%%%%%%%%%%%%%%%%%
%PAGE 19 PAGE 19 PAGE 19 PAGE 19 PAGE 19 PAGE 19 PAGE 19 PAGE 19 PAGE 19 PAGE 19
%%%%%%%%%%%%%%%%%%%%%%%%%%%%%%%%%%%%%%%%%%%%%%%%%%%%%%%%%%%%%%%%%%%%%%%%%%%%%
Just for curiosity, consider the \textit{general} solution of (\ref{eq37})
\begin{equation}\label{eq40}
\vec{v} = \alpha_+ \begin{pmatrix}
c_+ \\ s_+
\end{pmatrix} +\beta_+ \begin{pmatrix}
-s_+ \\ c_+
\end{pmatrix} + \alpha_- \begin{pmatrix}
c_- \\ s_-
\end{pmatrix}+\beta_- \begin{pmatrix}
-s_- \\ c_-
\end{pmatrix},
\end{equation}
and insert into (\ref{eq38}); by a tedious elementary calculation,
\begin{equation}\label{eq41}
\begin{split}
\vec{v}\,^2 - 1 & = \gamma_+^2 + \gamma_-^2 -1 + g(v) \\
\vec{v}'^2 & = (\gamma_0 +1)^2 \gamma_+^2 + (\gamma_0 - 1)^2 \gamma_-^2 + (\gamma_0^2 -1)g(v) \\
\vec{v} \times \vec{v}' & = (\gamma_0 +1)\gamma_+^2 + (\gamma_0 -1)\gamma_-^2 + \gamma_0 g(v) \\
\end{split}
\end{equation}
with $\gamma_{\pm} := (\alpha_{\pm}^2 + \beta_{\pm}^2)$, $g(v) := 2(cos(2v))(\alpha_+ \alpha_- + \beta_+ \beta_-)+ 2(sin(2v))(\alpha_+ \beta_- - \alpha_+ \beta_-)$,one finds that the $g^2(v)$ terms will match automatically, while the terms linear in $g(v)$ give
\begin{equation}\label{eq42}
(\gamma_0+1)\gamma_+^2 + (\gamma_0 - 1)\gamma_-^2 = \dfrac{1}{2}\gamma_0(\gamma_0^2 -1);
\end{equation}
using (\ref{eq42}) the constant terms imply a (still rather involved) quadratic equation for $\gamma_-^2$ (or $\gamma_+^2$) which, putting (\textit{either} $+$ \textit{or} $-$)
\begin{equation}\label{eq43}
\gamma_{\pm}^2 = \left(\dfrac{\gamma_0 \mp 1}{2}\right)^2 \cdot x
\end{equation}
has $x = 1$ as a double root, so that (from (\ref{eq42})) (\ref{eq43})$_{x=1}$ is the unique solution for $\gamma_{\pm}^2$,
%%%%%%%%%%%%%%%%%%%%%%%%%%%%%%%%%%%%%%%%%%%%%%%%%%%%%%%%%%%%%%%%%%%%%%%%%%%%%
%PAGE 20 PAGE 20 PAGE 20 PAGE 20 PAGE 20 PAGE 20 PAGE 20 PAGE 20 PAGE 20 PAGE 20
%%%%%%%%%%%%%%%%%%%%%%%%%%%%%%%%%%%%%%%%%%%%%%%%%%%%%%%%%%%%%%%%%%%%%%%%%%%%%
confirming the previously found solution (\ref{eq39}), as rotations, reflections, and shifts $v\rightarrow v_0$ leave $\gamma_{\pm}^2$ invariant. The more interesting observation, however, is that (\ref{eq38}) trivially follows from the 2 (\textit{not} reparametrization invariant) equations
\begin{equation}\label{eq44}
\vec{v}'^2 = (\gamma_0^2 - 1)(\vec{v}\,^2 - 1)
\end{equation}
\begin{equation}\label{eq45}
\vec{v} \times \vec{v}' = \gamma_0(\vec{v}\,^2 - 1)
\end{equation}
which not only trivially imply (via (\ref{eq41}))
\begin{equation}\label{eq46}
\gamma_{\pm}^2 = \dfrac{(\gamma_0 \mp 1)^2}{4},
\end{equation}
as the resulting linear equations (from (\ref{eq44}), resp. (\ref{eq45}))
\begin{equation}\label{eq47}
\begin{split}
2(\gamma_0 + 1)\gamma_+^2 - 2(\gamma_0 -1)\gamma_-^2 & = 1 - \gamma_0^2 \\
\gamma_+^2 - \gamma_-^2 & = -\gamma_0
\end{split}
\end{equation}
immediately give (\ref{eq46}); but \textit{also} imply (\ref{eq37}), by simple differentiation: $\vec{v}'\vec{v}'' = (\gamma_0^2 -1)\vec{v}\vec{v}'$ gives
\begin{equation}\label{eq48}
\vec{v}'' + (1-\gamma_0^2)\vec{v}' = \varepsilon A \vec{v}',
\end{equation}
while $2\gamma_0\vec{v}\vec{v}' = \vec{v} \times \vec{v}'' = -\vec{v}A\vec{v}''$ gives
\begin{equation}\label{eq49}
A\vec{v}'' + 2\gamma_0\vec{v}' = \delta A \vec{v},
\end{equation}
hence (determining the unknowns $\varepsilon$ and $\delta$)(\ref{eq37}).\\
%%%%%%%%%%%%%%%%%%%%%%%%%%%%%%%%%%%%%%%%%%%%%%%%%%%%%%%%%%%%%%%%%%%%%%%%%%%%%
%PAGE 21 PAGE 21 PAGE 21 PAGE 21 PAGE 21 PAGE 21 PAGE 21 PAGE 21 PAGE 21 PAGE 21
%%%%%%%%%%%%%%%%%%%%%%%%%%%%%%%%%%%%%%%%%%%%%%%%%%%%%%%%%%%%%%%%%%%%%%%%%%%%%
How could one have guessed this? (Just as generally with integrable systems - which the above argument definitely shares some flavor with - it is easy to check, but non-trivial to find) (\ref{eq44}) and (\ref{eq45}) are the conditions for the metric to be conformally constant, i.e. such that resulting from the condition that $G_{00} = \dot{x}^2 = 1-\vec{v\,^2}$, $G_{01} = \dot{x} x' = -\vec{v}\times \vec{v}' \frac{1}{\omega}$ and $G_{11} = -\frac{\vec{v}'^2}{\omega^2}$ are all proportional to the same function, with the  \textit{constant} proportionality factors read off from (\ref{eq37}), resp. the matrix in (\ref{eq33});
\begin{equation}\label{eq50}
\dfrac{G_{01}}{G_{00}} = \dfrac{\gamma_0}{\omega},\; \dfrac{G_{11}}{G_{00}} = \dfrac{\gamma_0^2 - 1}{\omega^2}.
\end{equation}
To directly check that (\ref{eq39}) satisfies (\ref{eq38}) is of course simple.

\end{document}